# Doping dependence of the anisotropic quasiparticle interference in NaFe$_{1-x}$Co$_x$As iron-based superconductors

Peng Cai[1,2], Wei Ruan[1,2], Xiaodong Zhou[1,2], Cun Ye[1,2], Aifeng Wang[3], Xianhui Chen[3], Dung-Hai Lee[4,5], and Yayu Wang[1,2] †

[1]*State Key Laboratory of Low Dimensional Quantum Physics, Department of Physics, Tsinghua University, Beijing 100084, P. R. China*

[2]*Collaborative Innovation Center of Quantum Matter, Beijing, China*

[3]*Hefei National Laboratory for Physical Science at Microscale and Department of Physics, University of Science and Technology of China, Hefei, Anhui 230026, P.R. China*

[4]*Department of Physics, University of California at Berkeley, Berkeley, CA 94720, USA*

[5]*Materials Science Division, Lawrence Berkeley National Lab, Berkeley, CA, 94720, USA*

† Email: yayuwang@tsinghua.edu.cn

We use scanning tunneling microscopy to investigate the doping dependence of quasiparticle interference (QPI) in NaFe$_{1-x}$Co$_x$As iron-based superconductors. The goal is to study the relation between nematic fluctuations and Cooper pairing. In the parent and underdoped compounds, where four-fold rotational symmetry is broken macroscopically, the QPI patterns reveal strong rotational anisotropy. At optimal doping, however, the QPI patterns are always four-fold symmetric. We argue this implies small nematic susceptibility and hence insignificant nematic fluctuation in optimally doped iron pnictides. Since $T_C$ is the highest this suggests nematic fluctuation is *not* a prerequistite for strong Cooper pairing.

Most unconventional superconductors possess a complex phase diagram in which certain exotic order is intertwined with the superconducting (SC) phase [1]. In the iron-based superconductors, such intertwined electronic orders are the antiferromagnetism and *nematicity*. The first evidence for a liquid-crystal-like electronic order was observed in the parent state of $Ca(Fe_{1-x}Co_x)_2As_2$ by scanning tunneling microscopy (STM) [2]. The QPI features in this compound show that the impurities induce a dimer-like deficit/excess of density of states (DOS). Moreover, the orientation of the dimer is aligned within the magnetic domains. Subsequently, the rotationally asymmetric electronic structure was directly mapped out by angle-resolved photoemission spectroscopy (ARPES) on detwinned crystals of $Ba(Fe_{1-x}Co_x)_2As_2$ [3], which also exhibit pronounced resistivity anisotropy within the FeAs plane [4, 5]. It was proposed that the electronic anisotropic impurity halos are an important cause of the transport anisotropy [6, 7].

The most important question concerning the electronic nematicity is its relation to superconductivity in the iron pnictides [8-16]. Many experimental probes have been applied to clarify this issue. In particular, Ref. [17] reports the "divergence of nematic susceptibility" near the composition where $T_C$ of $Ba(Fe_{1-x}Co_x)_2As_2$ is the highest, hinting the possible role nematic fluctuations play in Cooper pairing. Studying impurity induced change in electronic structure not only yields information on macroscopic symmetry breaking [18], but also on the severity of order parameter fluctuations when the symmetry is preserved *macroscopically* [16]. Recently a STM study of the parent compound of NaFeAs reveals even in the tetragonal phase the QPI patterns are rotationally anisotropic [19]. This is presumably induced by impurities and local strains in the presence of large nematic susceptibility. However, up to date there is no report on the doping dependence of QPI patterns in the iron pnictides, which is crucial for elucidating the relation between the electronic nematicity and superconductivity.

In this work we use QPI imaging STM to study the symmetry breaking and nematic fluctuations in $NaFe_{1-x}Co_xAs$ iron-based superconductors from the parent to optimally doped regime. The variation of the QPI patterns shows that the electronic structure becomes more isotropic with increasing doping. Most importantly, in the optimally doped compound we find no evidence of nematic fluctuations, which suggests nematic fluctuation does not play an important role in strong Cooper pairing in the iron pnictides.

Figure 1(a) displays the schematic phase diagram of the $NaFe_{1-x}Co_xAs$ system, in which the structural, magnetic, and SC transitions are marked by solid symbols. The parent NaFeAs has a structural transition from the high $T$ tetragonal to the low $T$ orthorhombic phase at $T_S$ = 50 K, followed by the formation of a stripe-like spin density wave (SDW) order at $T_{SDW}$ = 40 K. The underdoped sample ($x$ = 0.014) shows the structural, SDW and SC transitions at $T_S$ = 32 K, $T_{SDW}$ = 22 K and $T_C$ = 16 K, respectively. At optimal doping ($x$ = 0.028) the only phase transition is the SC transition at $T_C$ = 20 K, and the crystal remains tetragonal at all temperatures. Thus STM studies on these three samples allow us to track the anisotropic electronic structure from the orthorhombic/AFM parent phase to the tetragonal/SC phase via an intermediate coexistence phase [20-22]. All the STM results reported in this paper are measured at $T$ = 5 K.

We start the investigations from the parent compound. Fig. 1(c) displays the differential conductance ($dI/dV$) map measured at sample bias $V$ = -20 mV on cleaved NaFeAs crystal, which represents the spatial distribution of the electron DOS with energy $\varepsilon = -20$ meV relative to the Fermi energy ($E_F$). The dominant feature here is the existence of many identical dimer-shaped DOS depressions, as marked by the yellow ellipses. The long-axis of the dimers is rotated by 45 degree with respect to the square lattice of the surface Na layer measured on the same area [Fig. 1(c) inset]. Comparison to the schematic structure [Fig. 1(b)] reveals that the dimers are all aligned along a particular Fe-Fe bond direction (we cannot distinguish the orthorhombic *a* and *b* axes). The overall features bear strong resemblance to the unidirectional dimer patterns observed in the Ca(Fe$_{1-x}$Co$_x$)$_2$As$_2$ parent state [6]. Fig. 1e displays the Fourier transform (FT) of the *dI/dV* map, which reveals three parallel bars pointing to the $\Gamma-M$ (zone corner) direction of the folded Brillouin zone (BZ) shown in Fig. 1(d). The momentum space structure is also similar to that in Ca(Fe$_{1-x}$Co$_x$)$_2$As$_2$, although the quantitative values of the QPI wavevectors are slightly different.

Figures 2(a) to 2(d) show the *dI/dV* maps measured on the same field of view as Fig. 1(c) but with four different bias voltages. The spatial patterns show apparent variations with bias, but the main features are concentrated near the dimers. Although there is no Co doping in the parent NaFeAs studied here, there are plenty of Fe-site defects which may originate from vacancies or impurities [Fig. 1(c) inset]. Closer examination of the dimer patterns reveals that as the energy changes, the intra-dimer length also varies as indicated in the figures. Moreover, when the bias becomes positive [Fig. 2(c) and 2(d)], the dimers reverse from DOS suppression (dark pits) to enhancement (bright spots). The main features of the QPI results are consistent with that reported by another group in NaFeAs [19].

Next we turn to the underdoped compound. Figures 3(a) to 3(e) display the *dI/dV* maps taken on the underdoped sample with five different biases, which reveal a rather complex evolution. At $V = +12$ mV, there are apparent quasi-1D stripe-like interference patterns along one of the Fe-Fe bond directions in real space, demonstrating the lowering of the $S_4$ symmetry (namely 90° rotation plus the reflection about the Fe-plane) to $C_2$. The FT image in the inset also shows two bright spots along the blue arrow direction pointing to one of the BZ corner. This is caused by the quasi-periodic inter-stripe structure. Decreasing the bias down to $V = 0$ mV, the stripe-like patterns are still present and the distance between them remain unchanged, as can be seen by the constant QPI wavevector along the direction of the blue arrow. However, there are dispersive features along the stripes [can be seen more clearly in Fig. 3(b)], which are manifested by the two arc-like features along the red arrow direction in the FT. The distance between the arcs increases with decreasing energy. Upon further decrease of the bias to negative values, the distance between the arcs becomes comparable to that between the bright spots along the blue arrow. As a result the QPI patterns in real space become nearly square-like at $V = -12$ mV [Fig. 3(e)].

Figure 3f summarizes the dispersion of the above mentioned FT features as a function of bias for the underdoped samples. The position of the spots along the blue arrow is non-dispersive, indicating a constant inter-stripe distance of ~11($\pm$1) times the Fe-Fe bond length ($a_{Fe}$), which is quite different from 8 $a_{Fe}$ [2, 6] and 16 $a_{Fe}$ [23] reported previously in other systems. The separation between the arcs along the red arrow decreases smoothly with increasing bias and reaches ~0.17 $\pi/a_0$ at $E_F$. The electronic structure thus exhibits apparent $C_2$ symmetry, and the anisotropy becomes more pronounced at positive biases in the unoccupied states.

Last we come to the optimal doping. Figures 4(a) to 4(e) display the d$I$/d$V$ maps measured on the optimally doped sample at five different biases. The dispersive QPI features are more pronounced, and the dispersion is stronger than that in the underdoped sample. However, there is no evidence of the lowering of $S_4$ rotational symmetry in any of the maps. Instead the dominant QPI features are always square-like patterns. Both features are manifested clearly by the FT images shown in the insets, which reveal well-defined QPI spots along both the blue and red arrows. The position of the interference spots change rapidly with energy but the $S_4$ symmetry is preserved in all maps. Fig. 4(f) summarizes the dispersion relations along the two perpendicular directions, which are equivalent to each other as expected for a $S_4$-symmetric electronic structure. Interestingly, the dispersions show approximate particle-hole symmetry with respect to $E_F$. For negative bias the position and the dispersion of the QPI spots highly resemble that of the $h_3$ hole band in stoichiometric LiFeAs [24]. We note that the lack of nematic electronic order in optimally doped NaFe$_{1-x}$Co$_x$As is contrary to recent torque magnetometer measurement showing that the nematicity extends to the overdoped regime of BaFe$_2$(As$_{1-x}$P$_x$)$_2$, where the lattice retains the $S_4$ symmetry [25].

We next discuss the implications of the above QPI results in NaFe$_{1-x}$Co$_x$As. Due to the multiband nature of the electronic structure, it is technically difficult to obtain a quantitative understanding of the QPI patterns to the level of that achieved in the cuprates [26, 27]. Therefore, the main focus of the discussion here is on the symmetry, i.e., whether the FT-QPI images exhibit $C_2$ or $S_4$ rotational symmetry.

In the parent compound, the dominant features are randomly distributed, unidirectional dimer-like impurity states. As discussed in Ref. [6], the main effect of these local impurity states are to provide anisotropic scattering of the quasiparticles, which explains the resistivity anisotropy between the orthorhombic *a* and *b* axes. In the underdoped regime, the QPI shows strong $S_4$ symmetry breaking. The arc-like dispersion along the stripes and constant inter-stripe periodicity are characteristic features of the nematic stripy QPI patterns.

The most important finding of this work is that at optimal doping the electronic structure is $S_4$ symmetric. In view of the fact that impurity/local strain can induce local electronic anisotropy in the $S_4$ symmetric phase when the nematic susceptibility is large, and the fact that at optimal doping there is no evidence of any electronic anisotropy suggests that the nematic susceptibility is low at optimal doping. A low nematic susceptibility in turn means there is barely any nematic fluctuation. Since $T_C$ is the highest at optimal doping this can only mean one thing: nematic fluctuations are not important for strong Cooper pairing in the iron pnictides. This same conclusion was reached in a theoretical paper recently [28].

There are still a number of open questions. (1) It would be interesting to follow the evolution of the QPI patterns as the temperature is lowered to the tetragonal-orthorhombic structural phase transition from above. This will allow one to gain quantitative understanding of how increasing nematic susceptibility affects the QPI anisotropy. (2) It will be interesting to measure the QPI patterns above the superconducting transition at optimal doping to confirm that nematic fluctuations are weak in the *normal state*. (3) Previous structural studies on Ba(Fe,Co)$_2$As$_2$ indicate the weakening of orthorhombicity upon the superconducting transition [29]. It will be interesting to see if there is any change of the QPI anisotropy in our underdoped compound upon the superconducting transition. In order to resolve these issues, high resolution QPI measurements to elevated temperatures are required. These will be a series of highly challenging, but highly informative experiments that deserve future investigations.

In summary, STM-QPI studies in NaFe$_{1-x}$Co$_x$As reveal that the electronic structure of iron pnictides become more isotropic with increasing doping. In particular, we demonstrate unambiguously that the optimally doped sample has a S$_4$ symmetric electronic structure in the ground state, which suggests that nematic fluctuations are insignificant for the highest $T_C$. This suggests strong nematic fluctuations are not prerequisite for strong Cooper pairing in the iron pnictides.

This work was supported by the National Natural Science Foundation and MOST of China (grant No. 2010CB923003, 2011CBA00101, and 2012CB922002). D.H.L. was supported by DOE grant number DE-AC02-05CH11231.

Figure Captions:

FIG. 1 (color online). (a) Schematic phase diagram of $NaFe_{1-x}Co_xAs$. (b) Schematic top view of the lattice structure. The dashed green and solid blue squares show the ideal one-Fe ($a_{Fe-Fe}$ ~ 2.8 Å) and real two-Fe ($a_0$ ~ 4.0 Å) unit cells. (c) Differential conductance $dI/dV$($V$ = -20 mV) map acquired at $T$ = 5 K over a 710×710 Å$^2$ area on parent NaFeAs reveals dimer-shaped impurity states. All dimers lie along one of the Fe-Fe bond directions (white axes), determined by the comparison to the surface Na lattice (inset, 50×50 Å$^2$ area). (d, e) Fourier transform of conductance map in (c), showing three equally spaced bars aligned along the dimer direction. The solid blue square marks the folded BZ, and the solid blue circles indicate the Bragg peaks of the Na atoms.

FIG. 2 (color online). The $dI/dV$ maps measured at four different biases in the same area as Fig. 3(c). The yellow dashed ellipses mark the dimer-shaped impurities. Inset of (d): spatially averaged $dI/dV$ spectrum showing an asymmetric SDW gap in the parent state.

FIG. 3 (color online). (a)-(e) The $dI/dV$ maps acquired at 5 K over a 980×980 Å$^2$ area of the underdoped sample ($x$ = 0.014) at biases +12 mV (a), +6 mV (b), 0 mV (c), -6 mV (d), and -12 mV (e). Top right insets show the corresponding FT. The red ($q_1$) and blue ($q_2$) arrows in (a) are along the Fe-Fe bond directions. Bottom left inset in (a) displays atomically resolved topography, where the bright six-Na-atom rectangular pattern indicates the underlying Co atom that substitutes either of two Fe sites with equal probability. Bottom left inset of (e) is the spatially average $dI/dV$ spectrum, illustrating coexistence of SDW and SC in the underdoped regime [22]. (f), The ε vs q dispersion relation extracted from the FTs along the $q_1$ and $q_2$ directions.

FIG. 4 (color online). (a)-(e), The *dI/dV* maps and corresponding FTs (top right inset in each panel) measured at selected energies on a 800×800 Å$^2$ area of optimally doped NaFe$_{1-x}$Co$_x$As surface ($x$ = 0.028) at 5 K. Bottom left inset of (a) exposes the atomically resolved surface with more Co impurities. Bottom left inset of (e) is the spatially average *dI/dV* spectrum, revealing a particle-hole symmetric SC gap. (f), The $\varepsilon$ vs q dispersion relation along the q$_1$ (red) and q$_2$ (blue) directions indicated in (a). The dispersion shows no difference between the two Fe-Fe axes, indicating the absence of nematic order.

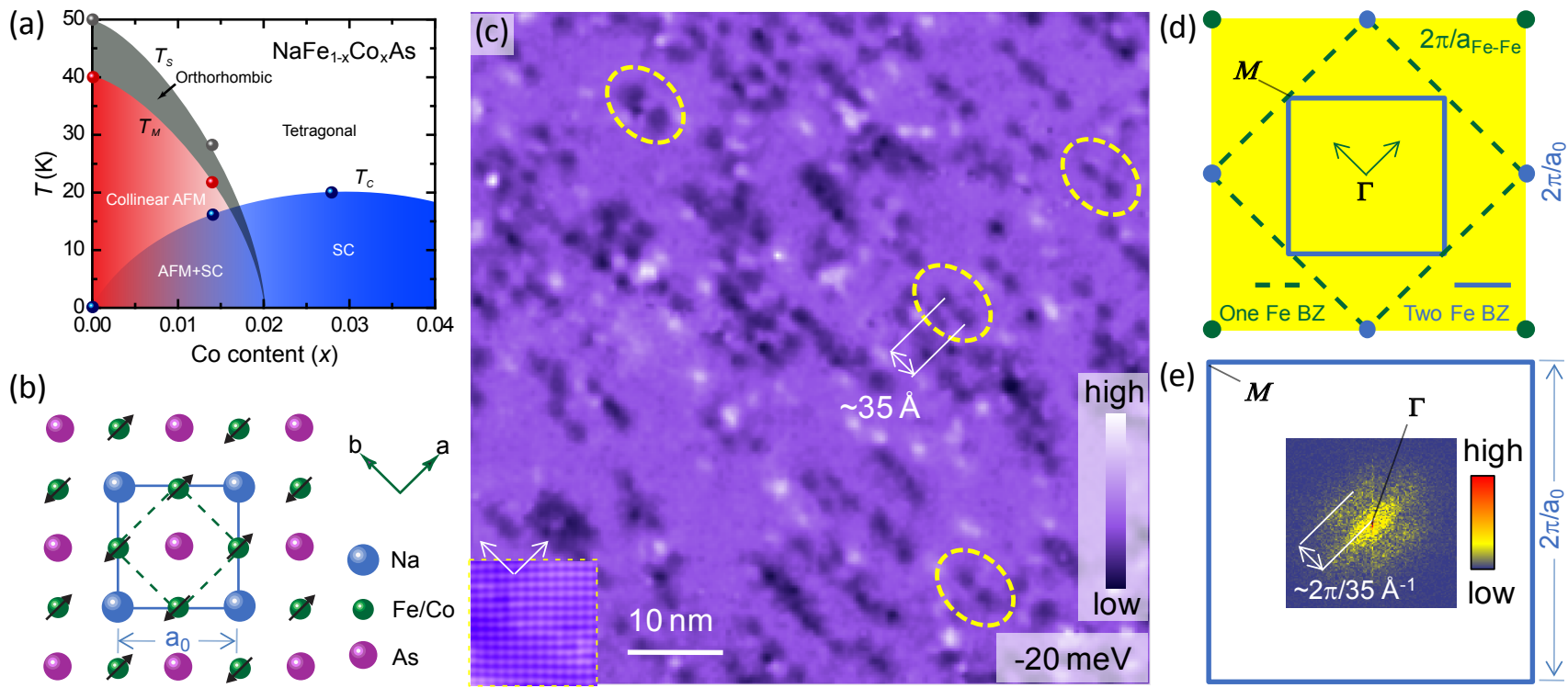

Figure 1

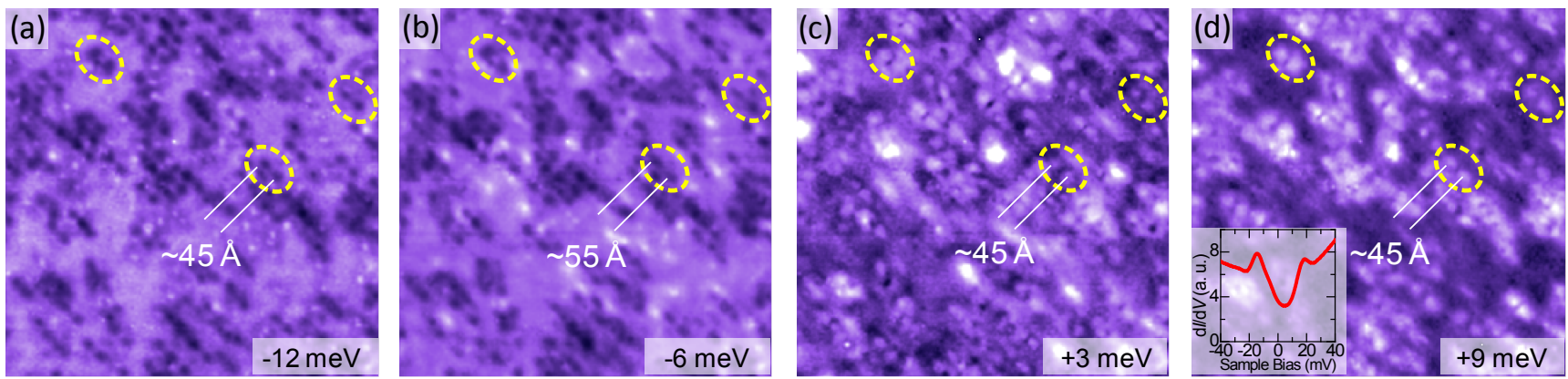

Figure 2

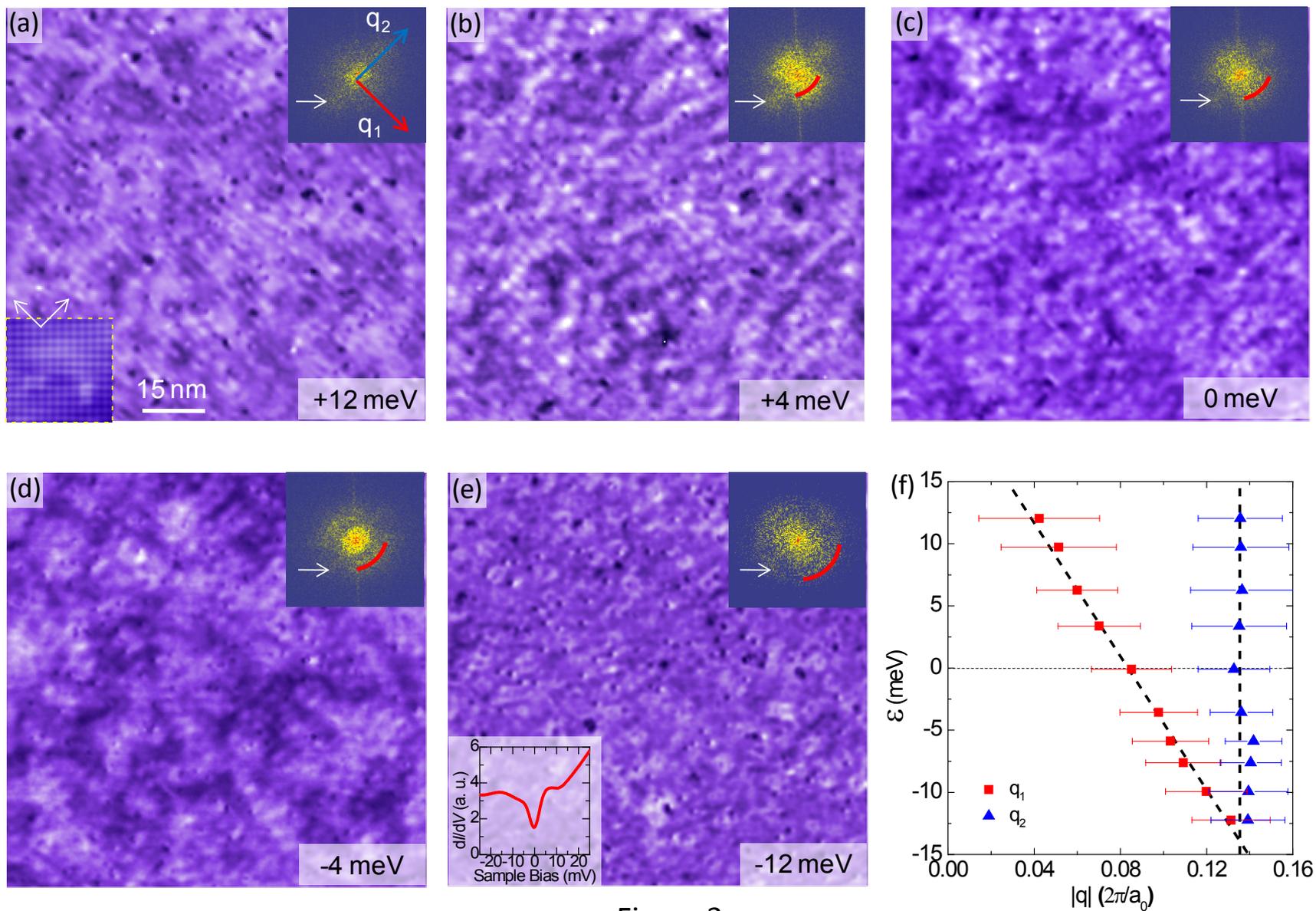

Figure 3

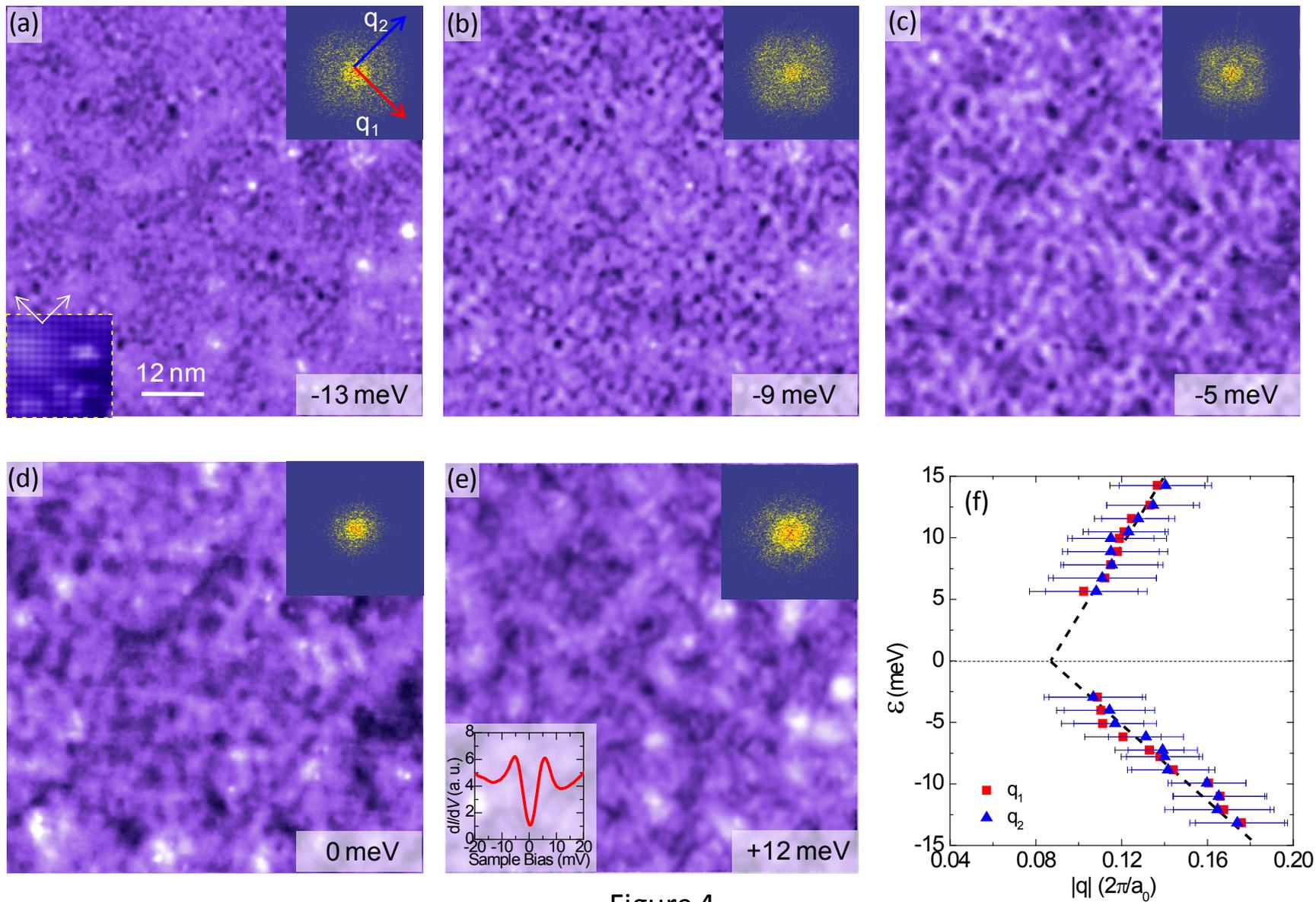

Figure 4